\documentstyle[12pt,epsf]{article}

\input{psfig}

\newcommand{\nc}{\newcommand}
\def\frac#1#2{{\textstyle {#1 \over #2}}}

\nc{\beq}{\begin{equation}}
\nc{\eeq}{\end{equation}}
\nc{\beqa}{\begin{eqnarray}}
\nc{\eeqa}{\end{eqnarray}}
\nc{\lsim}{\begin{array}{c}\,\sim\vspace{-21pt}\\< \end{array}}
\nc{\gsim}{\begin{array}{c}\sim\vspace{-21pt}\\> \end{array}}

\newcommand{\mysection}[1]{\setcounter{equation}{0}\section{#1}}

\begin{document}
\begin{titlepage}

\begin{center}

\vskip .5 in
{\large \bf  Spin Liquid Phases in  $2D$  Frustrated
XY Models}

\vskip .6 in
{
  {\bf  P. Simon}\footnote{simon@lpthe.jussieu.fr }
   \vskip 0.3 cm
   {\it   Laboratoire de Physique Th\'eorique et Hautes Energies}\footnote{Unit\'e associ\'ee au CNRS URA 280}\\
     {\it      Universit\'es Pierre et Marie Curie, Paris VI et Denis Diderot, Paris VII}\\
{\it  2 pl. Jussieu,
          75251 PARIS  cedex 05 }\\ 
}

  \vskip  1cm  
\end{center}
\vskip .5 in
\begin{abstract}
In this paper we consider the $J_1-J_2-J_3$ classical and quantum $2D$ XY model. Spin wave calculations show that a spin liquid phase still exists in the quantum 
case as for Heisenberg models. We formulate a semiclassical approach of these 
models based on spin wave action and use a variational method to study the role played by vortices.  Liquid and 
crystal phases of vortex could emerge in 
this description. These phases seem to be directly 
correlated with the spin liquid one and to its crystalline interpretation. 
\end{abstract}

\end{titlepage}

\renewcommand{\thepage}{\arabic{page}}
\setcounter{page}{1}
\mysection{Introduction  }
Low dimensional magnetic systems have been extensively studied these last years since they are  known to exhibit non trivial quantum behaviors. This comes in particular from the fact
that classical Heisenberg $2D$ anti-ferromagnets can order only at zero temperature  so that quantum fluctuations cannot {\it a priori} be neglected. In fact, it is now well-known that in  antiferromagnets, the quantum fluctuations can ($S < S_{critical}$) or cannot ($S> S_{critical}$) induce a transition at $T=0$ from a N\'eel ordered ground state to a quantum disordered one, characterized by short range order \cite{chn}. The enhancement of quantum fluctuations has naturally been looked for in frustrated spin systems. The combination of frustration and quantum fluctuations can possibly lead to a spin liquid state (for a review about spin liquid state see for example ref. \cite{schulz}). For the $2D$ Heisenberg model with nearest neighbor (NN) and next to nearest neighbors (NNN) antiferromagnetic interactions, Chandra and Dou\c{c}ot have shown by spin wave calculations that it  can be disordered at $T=0$ \cite{chan}. This has been confirmed by exact diagonalizations on a finite lattices \cite{dago,schulz}, by  series expansions \cite{singh}, and by a renormalization group analysis of the associated non linear sigma models \cite{suede,ioffe,ferrer}. The latter study has shown that the couplings of the models flow under renormalization group transformations towards a strong coupling regime when this liquid phase is approached \cite{ferrer}. Einarsson and Johannesson \cite{suede} have shown that precisely close to this liquid state, there is a proliferation of tological excitations in the path integral representation of the frustrated Heisenberg model \cite{hald}. They have suggested from these instanton considerations possible realizations of this disordered liquid phase in analogy with works of Sachdev {\it et al.} \cite{sachdev}.\\
In this article, we want to emphasize that there exists a system where it is likely that this relationship between topological excitations and this liquid state is much more direct, namely the quantum frustrated $2D$ XY model. We have indeed shown in a preceding paper that a $2D$ classical XY model with (NN) and (NNN) interactions has a point in the $({J_2\over J_1},T)$ plane, where $T_{KT}=0$ \cite{simon}. This result can be extended by adding a (NNNN) interaction with this time a whole line  of Kosterlitz-Thouless transitions at $T=0$, in the parameter space $({J_2\over J_1},{J_3\over J_1})$. This line is precisely  that around which, quantically, a spin liquid phase is found. We hypothesize in this article, that while quantum fluctuations can surely not be {\it a priori} neglected in this case, they are, together with vortices, responsible for a spin liquid state in this frustrated XY model.\\ 
\\\noindent 
The paper is organized as follows: In section 2, we consider the classical XY model with competing interactions.  The classical phase diagram at $T=0$ is recalled. It contains no spin liquid phase. Then, we  pay special attention to the role played by vortices and show that there is a critical line where  vortices are allowed  at $T=0$. This proves the need to incorporate quantum fluctuations. Indeed,  in section 3, we show that linear quantum spin wave computations predict  a spin
liquid phase (around this critical line) in the quantum frustrated XY model contrary to the classical model. Then, in section 4, we 
formulate  a general Ginzburg-Landau-Wilson (GLW)
semi-classical  action  from the quantum spin wave action in order to discuss the nature of this liquid phase. When approaching this phase, the spin stiffness, $\kappa(S)$ gets very small and we are {\it a priori} forced to take into account the effects of all quartic terms. We 
pay particular attention  to
  the non-perturbative sector of this action.
 We find a phase where pairs of vortices can be confined at a short finite lattice distance. The phase  described here, is some kind of liquid  of vortex. This liquid state is 
favored when spin waves become 
softer (when $\kappa(S)\to 0$) {\it i.e.} when the usual spin liquid phase is predicted! Finally,  section 5 contains a summary of the results and some concluding remarks.

\mysection{Classical frustrated XY models  }
The purpose of this section is to study the role played by vortices in classical frustrated
XY models, especially in the weakly frustrated phase.
Let us first consider the $2D$ classical $J_1-J_2$, XY model
\beq
\label{hamil}
H=-J_1\sum\limits_{<i,j>} cos(\theta_i-\theta_j)+J_2\sum\limits_{<<k,l>>} cos(\theta_k-\theta_l)~,
\eeq
with $J_1,J_2>0$ and $\theta_i$ the angles associated to the classical $O(2)$ spin $\vec{S}_i$. $<~,~>$ is for nearest neighbors (NN) and $<<~,~>>$ for next to nearest neighbors (NNN).
Two ground states 
are 
possible  in the model. When $|J_1|>2J_2$ the ground state is 
ferromagnetic, whereas, when  $|J_1|<2J_2$, the ground state consists  
of 
two independent diagonal sublattices with antiferromagnetic order \cite{henley}. \\
From a spin wave approximation, an effective action can be derived in the ferromagnetic phase as
\beq 
\label{sw1} 
{\cal A}_1 = {1\over 2T} \int d^2x   \left\{(J_1 -2J_2) (\nabla \theta)^2 + 
J_2(\nabla_x\nabla_y \theta)^2~\right\}.
\end{equation}
\noindent
The classical
vacuum is the standard ferromagnetic one, and therefore  the term $(\nabla_x\nabla_y 
\theta)^2$ is irrelevant according to standard perturbative arguments. In 
this 
case, the system can be well approximated by an XY model with an effective NN coupling 
constant $J_1-2J_2$, so that the action associated to a neutral pair of vortices is 
\beq
S_0= 4\pi( J_1-2J_2)\log ({\rho\over a}),
\eeq
 with $\rho$ 
 the distance separating the two vortices.  Therefore, there is a Kosterlitz-Thouless transition at the temperature $T_{KT}= {\pi\over 2}(J_1-2J_2)$. This result can be proved more rigorously by a Coulomb gas treatment of the original Hamiltonian (\ref{hamil}) \cite{simon}. It has to be noticed that the pairs of vortices play the role of the instantons in the non linear sigma model \cite{ogil}.
When  $J_1=2J_2$,   $T_{KT}=0$, a result which indicates that vortex solutions are allowed in the classical vacuum at $T=0$. This suggests that the model can have a non trivial behavior around this point when quantum fluctuations are considered.  When $J_1< 2J_2$, the behavior is drastically  different, the action  (\ref{sw1}) is somewhat  meaningless in this ``antiferromagnetic'' phase and  a new effective theory must be found \cite{simon}. Nevertheless, if we suppose that the notion of pair of vortices still makes sense in this case, the action associated to this pair  becomes negative so that it becomes  energetically favorable to fill the vacuum with pairs of vortices. Indeed, it has to be noticed that the antiferromagnetic 
ground state defined above (for  $J_1<{1\over 2 }J_2$) can be interpreted as a lattice of vortex-antivortex (for more details, see \cite{simon}).
 To summarize, we have seen that
 the  action associated to a pair of vortices indicates the  changes 
in the classical vacuum and especially that its value equals zero at the Lifshitz point . \\
\noindent
This analysis can be extended when a new coupling constant $J_3>0$, corresponding to a (NNNN) AF interaction, is added to the action (\ref{hamil}). When $J_1-2J_2-4J_3>0$, the 
classical vacuum is a standard ferromagnetic one. Let us consider the 
isotropic 
case $(J_2=2J_3)$ for convenience, the physics along the whole line $J_1-2J_2-4J_3=0$ being the same \cite{ferrer}. The associated action, again using a spin wave
approximation,  can be written as
\begin{equation}
\label{sw2}
{\cal A}_2 =\int d^2x~  {1\over 2T}\left\{A (\nabla \theta)^2 + 
 B(\nabla^2 \theta)^2~\right\},
\end{equation}
\noindent
where $A=(J_1-8J_3)$, $B=J_3$. In the  non-isotropic case, the extra-term
 $(J_2-2J_3)(\nabla_x\nabla_y\theta)^2$ has to be added to the action (\ref{sw2}) and does not change qualitatively the results.   
The saddle point equation is
\beq
\label{selle}
A \Delta\theta - B \Delta^2\theta=0~,
\eeq\
\noindent
with $\Delta=\nabla^2$ the Laplacian.
The neutral pair of vortices is a solution of (\ref{selle}) and its 
associated action reads $S_0= 4\pi( J_1-8J_3)\log ({\rho\over a})$. There is no dependence in $B$, so 
this 
result indicates that we are in a 
 situation similar to the previous model. Since $S_0=0$ 
at the critical line $J_1-2J_2-4J_3=0$, vortex
excitations are present at $T=0$ along the whole line suggesting that the quantum fluctuations have to be  included. Moreover, along this line, quadratic terms vanish. The propagator is then governed by quartic terms and so is of short range orde,r favoring a disordered state at $T=0$.
When 
$J_1-2J_2-4J_3<0$, one has to know which wave vector   minimizes the 
spin wave action (\ref{sw2}) in order to obtain the whole phase diagram. By this method, 
we obtain  the 
same phase diagram at the classical level as for Heisenberg spins (see 
\cite{joli}). This has been reported for completness  in Figure 1. At low $J_3$ 
and $J_2>{J_1\over 2}$, 
we recover the phase with independent AF order on each sublattices (noted 
AF$_2$) as in the $J_1-J_2$ model. The critical line separates the 
ferromagnetic phase from two helical incommensurate phases with respective 
wave vectors $(\pi,\pm Q_1), (\pm Q_1,\pi)$ (phase $C_1$) and $(\pm Q_2,\pm 
Q_2)$ (phase $C_2$), where $Q_1$, $Q_2$ are defined by 
$\cos(Q_1)={(2J_2-J_1)\over 4J_3}$ and $\cos(Q_2)={-J_1\over (2J_2+4J_3)}$ (see 
Figure 1). \\ \noindent
We have seen that vortex excitations are allowed on the classical critical line $J_1-2J_2-4J_3=0$.  Nevertheless, no liquid phase has 
been found around it. Consequently, frustration is not sufficient at the classical level to induce a spin liquid state and quantum fluctuations must be included.
 
\mysection{Linear spin wave theory in quantum frustrated XY models }

\noindent
We   consider now the 
quantum version of these models. We just present in this section  spin waves results that are well known  for the Heisenberg model \cite{chan,joli} but not to my knowledge for XY spins.  The Hamiltonian can be written as
\beq
H=\sum\limits_{i,j} J_{ij}(S_i^xS_j^x+S_i^yS_j^y)
\eeq
where $J_{ij}=J_1$ for a NN interaction, $J_{ij}=J_2$ for a NNN interaction and $J_{ij}=J_3$ for a NNNN interaction.  The zero
point fluctuations 
around different ordered states can be computed in the large  $S$ limit by different
methods: a standard one with 
Holstein-Primakov bosons (see for example \cite{joli}), and a second one with 
``a semi-polar'' representation 
of spin operators introduced by Villain \cite{vil1}. Both give the same results though the latter seems more appropriate for XY models.
Let us first consider the $J_1-J_2$, XY model (with $J_1<0, J_2>0$).\\
When $J_1>{1\over 2} J_2$ the ground state is ferromagnetic.
 The magnetization $<S^z>$ reads
\beq
<S^z>
= S+{1\over 2}-{1\over 8\pi^2}\int d^2k {2-\eta_k-\alpha_2(2-\eta'_k)\over \{
(2-\eta_k-\alpha_2(2-\eta'_k))^2-(\eta_k-\alpha_2\eta'_k)^2\}^{1\over 2} }.
\eeq
We have defined $\alpha_2={J_2\over J_1}$, $\eta_k={1\over 2}[\cos(k_x)+\cos(k_y)]$ and
$\eta'_k=\cos(k_x)\cos(k_y)$. The integral over $k$ runs over the first Brillouin zone $[-\pi,\pi]^2$ of the two dimensional lattice.
We see that the mean field value of $S^z$ is decreased by including the 
first corrections in ${1\over S}$.

\noindent
When $J_1<{1\over 2} J_2$ a collinear ground state is selected by spin waves \cite{henley}.
The same kind of calculation can be generalized, paying attention to the 
anisotropy, and leads to
\beq 
<S^z> = S+{1\over 2}-{1\over 8\pi^2}\int d^2k 
{|2-\cos(k_x)\cos(k_y)-\lambda_2(\cos(k_x)-\cos(k_y))|\over
2\{1-\cos(k_x)\cos(k_y)- \lambda_2(\cos(k_x)-\cos(k_y))\}^{1\over 2} },
\eeq  
where $\lambda_2={|J_1|\over 2J_2}$.\\
In Figure 2, we   represent spin wave  corrections as a  function of $J_2$. We 
can draw conclusions analogous to those found in the study of Heisenberg spins 
\cite{chan,joli}, 
namely linear spin wave theory (LSWT)  predicts at any $S$ a finite region 
around  the Lifshitz point where  the ground state is disordered (see Figure 
2). This seems to indicate the presence of a spin liquid state. 

Furthermore, 
the
quantum fluctuations diverge when $\alpha_2={J_2 \over |J_1|}\rightarrow 
{1\over 2}$ as
\beqa
<S^z>&\sim & S+{1\over 2}-\gamma\log^2(1-2\alpha_2)~,\alpha_2<{1\over 
2}\nonumber\\
<S^z>&\sim & S+{1\over 2} -\gamma'(2\alpha_2-1)^{{-1\over 2}}~,\alpha_2>{1\over 
2},
\eeqa
where $\gamma$ and $\gamma'$ are two unimportant real numbers.\\
\noindent
We find asymptotic behaviors  similar to Heisenberg spins. This is not in fact so
surprising since the Lifshitz point is the same. The validity of this first order approximation compared to other methods like Schwinger bosons mean field theory has been discussed in ref. \cite{ferrer}. Second order spin waves calculations go  along the 
same line and confirm the first order calculations (it also leads to a 
renormalization of the spin stiffness). Moreover, some exact diagonalizations 
on finite lattices agree qualitatively with  the existence a spin liquid state around the Lifshitz point\cite{dago,schulz}.\\ 
\noindent
A similar analysis can be performed for the $J_1-J_2-J_3$, XY model.
We  are interested in the region around the classical critical line 
$J_1-2J_2-4J_3=0$ which separates the ferromagnetic 
phase from two helical phases as we 
have seen in section 2. The Villain method can be easily generalized for chiral 
phases 
\cite{vil1}.
The pitch wave vector $\vec{Q}_{0,cl}$
 is defined by $\partial_Q J_{Q_0,cl}=0$ with $J_Q$ the Fourier transform 
associated to $J_{ij}$. The spin-wave frequency and the staggered magnetization
are in this case
\beqa
\label{spw}
\omega_k&=&2\sqrt{J(Q_0)[J(Q_0)-{J(Q_0+k)+J(Q_0-k)\over2}]}\\
\label{spw1}
<S^z>&=&S+{1\over 2}-{1\over 8\pi^2}\int d^2k{|-J(Q_0)+{J(Q_0+k)+J(Q_0-k)\over 
4}|\over
\sqrt{J(Q_0)[J(Q_0)-{J(Q_0+k)+J(Q_0-k)\over 2}]}}.
\eeqa
\noindent
Along the critical line $\kappa_{cl}=J_1 -2J_2-4J_3 $, the staggered 
magnetization has the same kind of divergences as in the $J_1-J_2$ model. For 
example, we evaluate at the point $J_2=0$ the leading divergence 
\beq
<S^z>\sim S+{1\over 2}-\alpha\log{(J_1-4J_3 )}.
\eeq
\noindent
This behavioris similar for   the whole critical line (apart from the 
Lifshitz point). We can notice that we recover directly spin wave calculations in the ferromagnetic phase by taking  $Q_0=0$.
It supports the fact that we can go
continuously from the ferromagnetic phase  to the chiral phases as it is 
already the case classically. These states have the same symmetries contrary to 
the case of the ferromagnetic and collinear phases separated by a Lifshitz 
point, and thus both states can be described by eqs. (\ref{spw}),(\ref{spw1}). 
\noindent
Since results for XY and Heisenberg spins are very similar at this order, it
is reasonable to think that higher 
order corrections will have similar effects like a renormalization of the 
spin stiffness $\kappa_R(S)$. It can be used to define the quantum critical line 
as $\kappa_R(S)=0$ \cite{ferrer}. Therefore, to summarize this section, we have shown that  spin waves predict the presence of a spin liquid phase around $\kappa_R(S)=0$ generated by
the combination of quantum fluctuations and frustration.

\mysection{A semi-classical treatment of quantum frustrated 
 XY models}
 
\noindent
In this section, we want to find an effective action able to describe the quantum behavior of the $2D$, $J_1-J_2-J_3$, XY model especially near the critical line.
The most rigorous way to find an effective action for spin models is to 
  decompose the pure spin Hamiltonian on a basis of coherent states, which 
enables
a path integral description of a single spin \cite{hald,suede}. Another 
possibility is to formulate directly the most general 
effective action from the spin wave action and symmetry considerations. The latter strategy has been already used by Ferrer to study the scaling 
properties of the $2D$  $J_1-J_2-J_3$ quantum
Heisenberg model \cite{ferrer}. We  follow a similar strategy. \\In our case, we have seen from the spin wave analysis
that quartic terms coming from NNN and NNNN interactions have to be included near the
critical line since quadratic terms vanish on this line. Therefore, the spin waves have a propagator of the following form ${\cal P}_{SW}^{-1}=a~\vec{k}^2 + b~\vec{k}^4$ (in the isotropic case).
Following 
Amit {\it et al} \cite{amit}, we wonder what physical systems could be described by a theory having the anomalous propagator $k^2+k^4$ and how it could be relevant for the understanding of the quantum phase diagram of the $2D$ $J_1-J_2-J_3$ XY model. Therefore, we are interested in the sequelin  the most  general Landau-Ginzburg-Wilson (LGW) action for an XY-spin 
$\vec{S}$, with such an anomalous propagator
\beq
\label{action}
{\cal A}= \int d^2x\left[A|\nabla\vec{S}|^2 + B|\nabla^2\vec{S}|^2 + B'|\nabla 
\vec{S}|^4 + {\cal V}(\vec{S})\right]~,
\eeq 
\noindent
where ${\cal V}(\vec{S})=r_0|\vec{S}|^2 +\lambda |\vec{S}|^4$. This effective action has a purely classical origin. Indeed, $A=(J_1-2J_2-4J_3)$, $B=J_3$ and $B'=0$ corresponds to the classical spin wave action (\ref{sw2}). We will now make an apparently ``crude '' approximation, namely we suppose that the quantum action has a form similar to (\ref{action}) except an extra-dimension and a renormalization of coupling constants, such that the values given above for $A,B,B'$ are no longer valid. Therefore,  the quantum model takes {\it a priori} different values in the  space of coupling constants. \\
Nevertheless, we have an important constraint from spin wave calculations: the quantum critical line is always characterized by $A\sim\kappa_r(S)\to 0$. Under the above hypothesis, this implies that the main differences between the classical and quantum situations rely in the behavior of higher derivative terms. It has to be noticed that similar effective actions have been derived in references \cite{ioffe,ferrer} for frustrated Heisenberg models.
  We do not claim that this simple effective action contains all the physics associated with the spin liquid phase but nevertheless, we will see that it can be useful to understand the role played by instantonic sector when the critical line corresponding to $A\sim\kappa(S)\to 0$ is approached. \\
 We suppose the spin $S$ to be large and analyze phase fluctuations as it is common for XY spins. Therefore, we write
$S(r)=S_0e^{i\theta(r)}$ where $S_0$ is 
spatially uniform and depends on the parameters of the model (see spin wave calculations). In fact, it just corresponds to the large $S$ version of the 
Villain semi-polar representation \cite{vil1}. Fluctuations around $S_0$ will be discussed 
later in the article. In this case, the action (\ref{action})  reads
\beq
\label{action1}
{\cal A}= \int d^2x\left[A S_0^2 (\nabla\theta)^2 + B S_0^2a^2(\nabla^2\theta)^2 
+  C S_0^2 a^2(\nabla \theta)^4  \right]~,
\eeq 
\noindent
with $C=B+B'S_0^2$. The lattice spacing $a$ is introduced in order to have 
dimensionless coupling constants. The potential part has been omitted because the main effects we will study involve the derivative part. Normally, the operator $|\partial_{\mu}\vec{S}\partial_{\nu}\vec{S}|^2$ should have been included in original action (\ref{action}) but, since it gives a similar contribution as $|\nabla\vec{S}|^4$, it has also been omitted.\\\noindent 
 To study this action, it seems useless to use perturbative arguments,
because, first the critical line ($A\sim\kappa(S)\to 0$) is characterized by a strong coupling regime 
\cite{ferrer}, and second the higher operators  are irrelevant 
according to the usual perturbation scheme. 
Because of the non linear term $(\nabla \theta)^4$, it is difficult to find the 
saddle points  analytically. Nevertheless, in section 2, we have noticed 
that the action associated to a  vortex-antivortex pair, characterized by 
the  
relative distance $\rho$, indicates the changes of the ground state 
when approaching the critical line (moreover, it is a genuine solution when $C=0$). Therefore, we use a variational approach based on a  wave function of pairs of vortices  defined by
\beq
\theta_0=\arctan{{y-y_1\over x-x_1}}- \arctan{{y-y_2\over x-x_2}}~,
\eeq
\noindent
with $\vec{r_1}=(x_1,y_1)$, $\vec{r_2}=(x_2,y_2)$ the position of both vortices
($\rho^2=|\vec{r_1}-\vec{r_2}|^2$). Introducing $\theta_0$ in (\ref{action1}),
we  have to compute 
\beq
\label{inte}
I=\int\int d r~ (\nabla \theta_0)^4=\int\int d^2 \vec{r} {\rho^4 a^2 
\over 
|\vec{r}-\vec{r_1}|^4|\vec{r}-\vec{r_2}|^4}.
\eeq
Of course, we must  regularize the integrals. The lattice constant $a$ is
the most natural regulator in our case. The computation is made in the 
appendix. We obtain the following action for the pair of vortices      
\beq
\label{vaction}
{\cal A}_V(\rho)= 4\pi S_0^2  \left[ A\log({\rho\over a }) + 
  C  \left(4\log({\rho\over a }){a^2\over \rho^2} 
-\alpha_2{a^2\over \rho^2}\right)
\right],  
\eeq
\noindent
with $\alpha_2> 0$ a  real number. When $A>0$ and $C>0$, this action 
is minimum for $\rho=a$ (our cut-off). Namely, we have a standard attractive 
increasing potential and charges tend to form dipoles and so does not contribute. 
The classical situation corresponds to $A=\kappa_{clas}>0$ $C=J_3>0$, hence in the weakly frustrated region, the vortices do not contribute except on the classical critical line.  The above conclusions are not altered in the anisotropic case. Nevertheless, when $C/A<C_0/A<0$, the 
situation becomes totally different. We can see on Figure 3, that  the potential 
has now a non trivial minimum $\rho_0$. This corresponds to a liquid phase of vortices. Moreover, when $C/A<C_1/A<C_0/A$ the action 
associated to the pair of vortices can be negative, meaning that it becomes 
energetically favorable to fill the semi-classical vacuum with   pairs of 
vortices separated by a size  $\rho_0$ of order $2a$ (see Figure 3). This result does not hardly  depend on the ratio $C/A$. It corresponds to a crystalline phase of vortices. Note that  the possible role of some higher order derivative terms in the non-perturbative sector has been already studied  in lattice gauge theories in refs \cite{pol} where some rather similar conclusions have been drawn. Before commenting these results, let us first perform one loop 
calculations, in order to see the contributions of such vortex effects to the 
path 
integral.\\
\noindent
In this perspective, we use the semi-classical aproximation and expand the field $\theta$ around $\theta_0$ as $\theta=\theta_0+\eta$ and  keep terms at most quadratic 
 in $\eta$ in the action (\ref{action}). The partition function reads
\beq
\label{fpart}
Z=\int {\cal D}\eta \exp(-S_0(\rho))\exp\left(-\int dx [~ A(\nabla\eta)^2+ 
B(\nabla^2\eta)^2 + 2C(\nabla \eta)^2(\nabla\theta_0)^2]\right).
\eeq
\noindent
The action in terms of $\eta$ after integration by parts reads
\beq
\label{etaaction}
{\cal A}(\eta)= 
A\eta\Box\eta-B\eta\Box^2\eta+2C\eta~[ (\nabla\theta_0)^2 
\Box-(\partial_{\mu}\partial^{\nu}\theta_0)^2]~\eta~.
\eeq
\noindent
After the gaussian integration, it remains to compute the determinant 
associated to the differential operator 
\beq
{\cal 
O}=A\Box-B\Box^2+2C\Box(\nabla\theta_0)^2-2C(\partial_{\mu}\partial^{\nu}\theta
_0)^2~.
\eeq
\noindent
Of course, it seems impossible to obtain the spectrum associated to this 
operator. Nevertheless, we can use the fact that  
$(\partial_{\mu}\partial^{\nu}\theta_0)^2$ and $(\nabla\theta_0)^2$ are 
ultralocal functions of $\vec{r}$ ($\rho_0\sim a$) whose  asymptotic behaviors are 
respectively in ${1\over r^6}$ and ${1\over r^4}$. Therefore, one can 
approximate the fluctuations determinant by $\det[A\Box-B\Box^2]$, which  just 
corresponds to the determinant associated to quantum spin waves. To be more 
explicit, we write 
\beq
\det[{\cal O}]=det[{\cal P}]\det[ 1-{\cal 
P}^{-1}(-2C(\partial_{\mu}\partial^{\nu}\theta_0)^2], 
\eeq
\noindent
 with  ${\cal 
P}=A\Box-B\Box^2+2C\Box(\nabla\theta_0)^2$. The second determinant can be 
treated perturbatively.  Indeed, asymptotically $$ {\cal 
P}^{-1}(-2C(\partial_{\mu}\partial^{\nu}\theta_0)^2(\vec{r})\sim\int d^2 
\vec{r_1} {\log|\vec{r}-\vec{r_1}|\over r_1^6},$$  and so does not contribute. 
A similar work can be done 
on $\cal{P}$ using the ultralocality of $(\nabla\theta_0)^2$ justifying the 
approximation\footnote{The treatment of {\cal P} is equivalent to solving a 
Schr\"odinger equation in a regularized attractive potential in ${1\over r^4}$. There will 
be just a few bound states plus the continuum spectrum similar to the case 
$C=0$
\cite{mes}.}. 
In the   partition function, we have now to integrate over $\rho$, $\rho$ 
playing now the role of the collective coordinate \cite{raj}. The only scale 
and 
translationally 
invariant measure is $d^2r d\rho \rho^{-3}$ \cite{polya}. In that case, the 
partition 
function reads
\beq
\label{fpart1}
Z=V \int\limits_0^{+\infty}  {d\rho\over \rho^3}e^{-S_0(\rho)}[\det{\cal 
O}(\rho) ]^{{-1\over 2}}~,
\eeq
\noindent  
where $a$ is the lattice cut-off. An interesting quantity often used is the 
ratio of the partition function in the unit and zero winding number. In fact, 
this ratio measures the weight of singular solutions in the path integral 
compared to 
spin waves. It will be clearly dominated by the tree sector, in so far as 
$\det[{\cal O}(\rho)]\sim \det[A\Box-B\Box]$. \\\noindent The last integral 
defined in 
(\ref{fpart1}) is carried out in the saddle point approximation around 
$\rho=\rho_0$. It yields  
\beq
Z=V e^{-S_0(\rho_0)} \int\limits_0^{+\infty} {d\rho\over \rho^2}e^{-2\pi S_0^2 
g(A,C,\rho_0)({1\over \rho^2}-{1\over \rho_0^2})^2}~[\det{\cal O}(\rho_0) 
]^{-{1\over 2}}~,
\eeq
\noindent
where $g(A,C,\rho_0) $ is a positive function defined by the second derivative
of the vortex action at $\rho=\rho_0 $. By performing the gaussian integration 
over $\rho$, the result is
\beq
Z =\lambda V e^{-S_0(\rho_0)}[\det{\cal O}(\rho_0) 2\pi 
S_0^2g(A,C,\rho_0) ]^{{-1\over 2}}, 
\eeq
\noindent
where $\lambda$ is an unimportant numerical constant. This expression goes one 
step further that the vortex action (\ref{vaction}). We find a 
competition between the 
exponential tree level factor and the fluctuation  determinant representing 
quantum spin wave effects that are long range ordered.\\ We are now obliged to wonder whether and
when this variational approach makes sense. It is clear that in the weakly 
frustrated phase, far from the quantum critical line, spin waves  
dominate  the path integral. In this region, the studies based on quantum spin 
wave calculations, omitting instanton configurations, have proved to capture 
the essential of the infrared behavior \cite{chan,suede,ferrer}. Nevertheless, 
when we are close 
to the quantum critical line, when $A$ gets very small compared to $B$ or $C$, spin 
waves become   softer and weaker and short range interactions (the quartic terms) are enhanced compared to spin waves. In that case, our variational method based on  topological defects
can be applied.
Moreover, as it was already mentioned, the approach of the quantum critical 
line is characterized by a strong coupling regime where topological excitations 
proliferate \cite{suede}. \\
\noindent
Let us now summarize and comment the results obtained so far. We have found that when spin waves fall down close to the quantum critical line, there is a range of parameters in the action (\ref{action1}) where pairs of vortices can proliferate and stabilize in a liquid phase or a crystal phase of vortices. It has to be noticed that the liquid phase of vortices corresponds to a small area in the parameter space. When  $C/A<C_1/A<0$, we found a crystalline phase where pairs of vortices are separated by a distance of order $2a$ (independently of the ratio $C/A$). Such a crystalline ground state is represented in Figure 4. This scenario supposes that in the quantum situation, there is a strong renormalization of the coupling constants able to change the classical behavior. This is our main hypothesis. We can not prove it, but such a possibility can not be ruled out  {\it a priori}. Moreover, the scenario we describe seems {\it a posteriori} consistent with the proliferation of vortices able to induce a crystalline ground state discussed by Einarsson {\it et al.}~.  
\\\noindent
Before concluding, a few remarks are in order. In our analysis, the higher order term $(\nabla\theta)^4$ plays an important role.
Amit {\it et al} \cite{amit} have already wondered about the role of such 
``dangerous irrelevant operators'' (see also ref. \cite{amit1}). In our study, we have shown   that its importance relies 
essentially in the non-perturbative sector. In the usual perturbation scheme, 
power counting arguments eliminate this kind of operators because they are 
irrelevant in the infrared limit. Nevertheless, the spin liquid phase is 
characterized by a strong coupling regime where   renormalization group 
technics fail. So, usual arguments used to eliminate such operators can not be  applied here. 
Near the critical line, the semi-classical vacuum becomes  disordered and dominated by short range 
order operators (higher gradient ones). In that case, it is not so surprising that such an operator 
can play a role. For more justifications, we refer the reader to ref \cite{pol} where this question is largely adressed.\\
In our analysis, the operator  $(\nabla\theta)^4$ has emerged because 
we have decomposed
$\vec{S}(x)=S_0 e^{i\theta(x)}$. If we consider amplitude fluctuations, namely if we  write $\vec{S}(x)=(\vec{S}_0+\Delta 
\vec{S'}(x))
e^{i\theta(x)}$ and integrate out the fluctuations of the order parameter 
( in a potential) we would generate series of operators like 
$(\partial_{\mu}\vec{S}\partial_{\nu}\vec{S})^n$ as described 
in \cite{chak}. We hope that these higher corrections do not change 
qualitatively  
 the physics presented in this section. Moreover, we will always find a range of parameters
where the scenario we have described should apply.

\mysection{Conclusion}
\vskip .3 in
In this paper, we have studied the $2D$ classical and quantum $J_1-J_2-J_3$, XY model on a 
square lattice. We have first shown that there is a classical critical line where vortices are present even at $T=0$. It justifies why quantum  fluctuations have to be included near this critical line. Quantum spin wave calculations predicts a spin liquid phase around this critical line. To study  the role played by vortices when approaching this spin liquid phase, we have considered a general GLW action deduced from spin wave calculations. We have found non-perturbatively a range of parameters where a crystal phase of vortices can take place around this critical line. This phase seems to be directly correlated with the spin liquid phase predicted by spin waves calculations. It is difficult to compare this crystalline phase with the possible disordered ground states of Heisenberg spins proposed by sachdev {\it et al.} \cite{sachdev} except that both descriptions have a short range crystalline order. Yet, this study has the advantage to show qualitatively how a non trivial liquid phase can emerge non perturbatively in two
dimensions at $T=0$. It should be very interesting  to investigate numerically this model. To test the validity of the scenario described in this paper, a possibility could be to look at the nature of the transition between the spin liquid phase and the chiral phases. It may correspond to a melting of this crystal of vortex phase as in ref. \cite{gabay}, induced by chiral spin waves. In that case, the transition could be of KT type.\\
Finally, this analysis
suggests that we may build classical spin models that can also be described by
an effective action similar to (\ref{action}). Such spin models would include 
multibody interactions. A general study of $3D$ classical spin systems without 
long-range order has been done by Alcaraz {\it et al} (\cite{savit} and 
references therein). They have done a general classification of these spin systems from symmetry considerations. This kind of models (especially their strange symmetry) could be useful to the description of 
certain aspects of disordered phases \cite{savit} in statistical systems. The link between 
such classical models and spin liquid phases is not  clear and will be 
the subject of future work.

 \vskip .8cm
{\bf  Acknowledgements.}

\vskip .3cm
I would  particularly like to thank  B. Delamotte and   B. Dou\c{c}ot  for useful suggestions 
and stimulating discussions. I also acknowledge D. Mouhanna, J. Richert for their comments 
about the manuscript.
 
\vskip .8cm
  {\bf  Appendix. } 
\vskip .3 in
In this appendix, we give the main steps of the computation of the integral 
(\ref{inte}). We place one vortex charge at the center $O(0,0)$ and the other 
in $A(\rho,0)$. In polar coordinate, the integral $I$ (\ref{inte}) reads
$$ 
I=\int r dr d\theta { \rho^4 a^2\over r^4(r^2+\rho^2-2r\rho\cos(\theta))^2}$$   
$$ 
=\int 2\pi dr {(r^2+\rho^2)sign(r^2-\rho^2)\over 
r^3(r-\rho)^3(r+\rho)^3}\eqno{(A1)}$$
The integration  of this rational function over $r$ gives
$$I=2\pi a^2\left[ \left(-4{\log({r\over a})\over \rho^2}+{1\over 2r^2} +  
2{\log({|r+\rho|\over a})\over \rho^2}-{7\over 
8\rho(r+\rho )}-{1\over 8(r+\rho)^2} + \right.\right.$$ 
$$\left.\left.+2{\log({|r-\rho|\over a})\over \rho^2}-{7\over 8\rho(r-\rho 
)}-{1\over 8(r-\rho)^2}\right)sign(r^2-\rho^2)\right]_0^{+\infty}\eqno{(A2)}$$
The contribution at $r=0$ and $r=\rho$ must be taken with our lattice 
regularization. We finally obtain
$$I=4\pi a^2\left[{4\log({r\over a})\over 
\rho^2}-{\alpha\over \rho^2}\right]\eqno{(A3)}$$
with $\alpha=2\log2+{5\over 32}$.
Notice that with this method, we recover the known result of $\int dx (\nabla 
\theta_0)^2=4\pi\log({\rho\over a})$ ( where $\theta_0$ represents a pair of 
vortices separated by a distance $\rho$).

\eject

\eject
\begin{center}
{\bf FIGURE CAPTIONS}
\end{center}

\vskip 0.5 truecm
FIG.1 The classical phase diagram for the $J_1-J_2-J_3$ XY model on a square 
lattice. $F$ corresponds to a ferromagnetic ground state, $AF_2$ to two 
decoupled sublattices with independent AF order and $C_1$, $C_2$ two 
incomensurate chiral phases.\\

\vskip 1. truecm
FIG.2 First quantum corrections to the lattice magnetization  in the $J_1-J_2$ 
XY model. The $O(1/S)$ spin wave theory predicts an intermediary region where 
the ground state is non-magnetic and thus can be a spin liquid phase.

\vskip 1. truecm
FIG.3 The action associated to a pair of vortex separated by a distance $\rho$.
Three cases have been represented: a positive action without any extremum, a 
positive action with a minimum in $\rho=\rho_0$ and a negative action with a 
similar minimum around $\rho=2a$.

\vskip 1. truecm 
FIG.4 A schematic representation of a lattice crystal of vortex separated by 
$\rho=2a$.
\eject

FIGURE 1

\begin{figure}
\epsfxsize=14cm
$$
\epsfbox{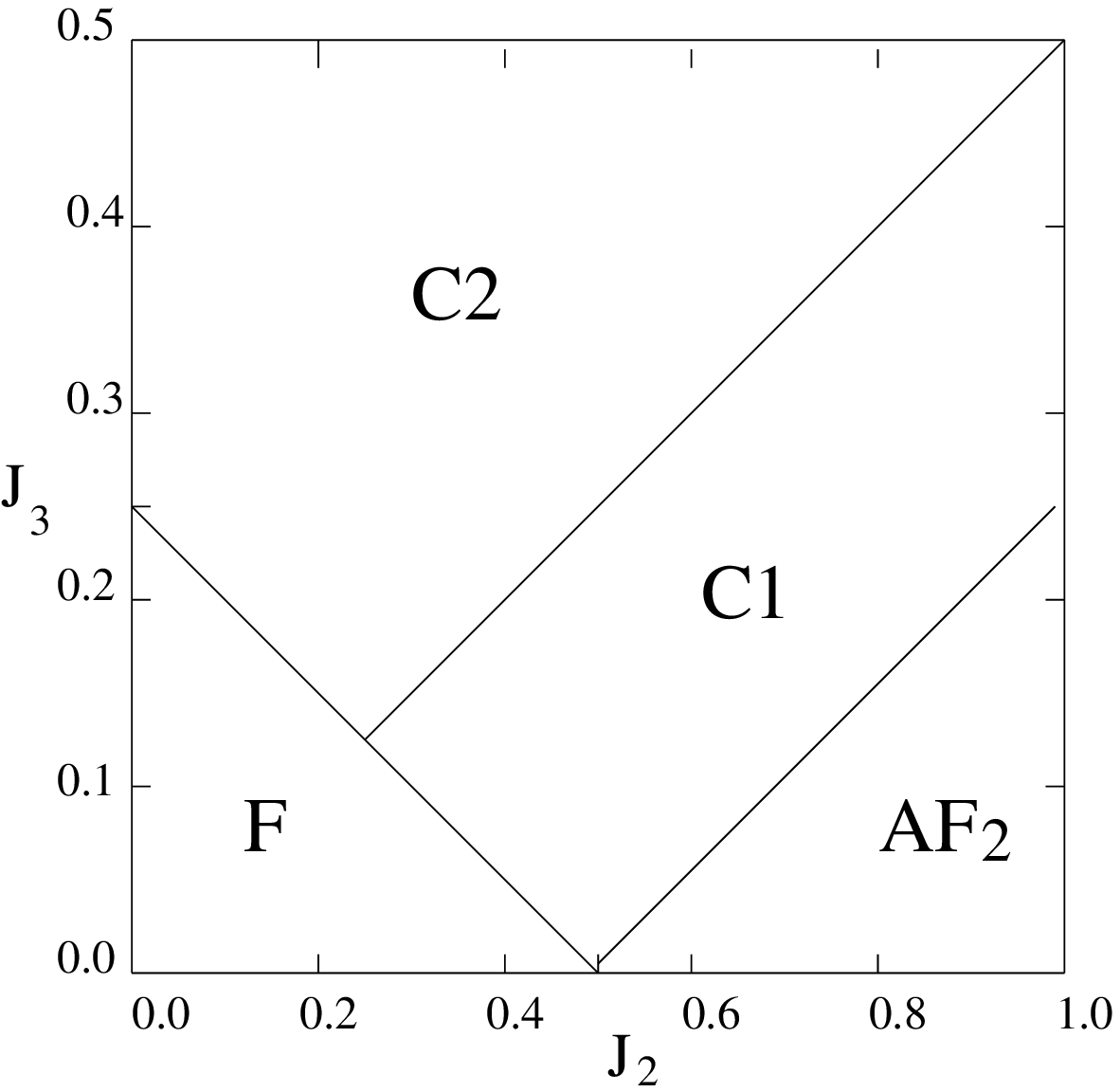}
$$
\end{figure}
\eject

\psfig{figure=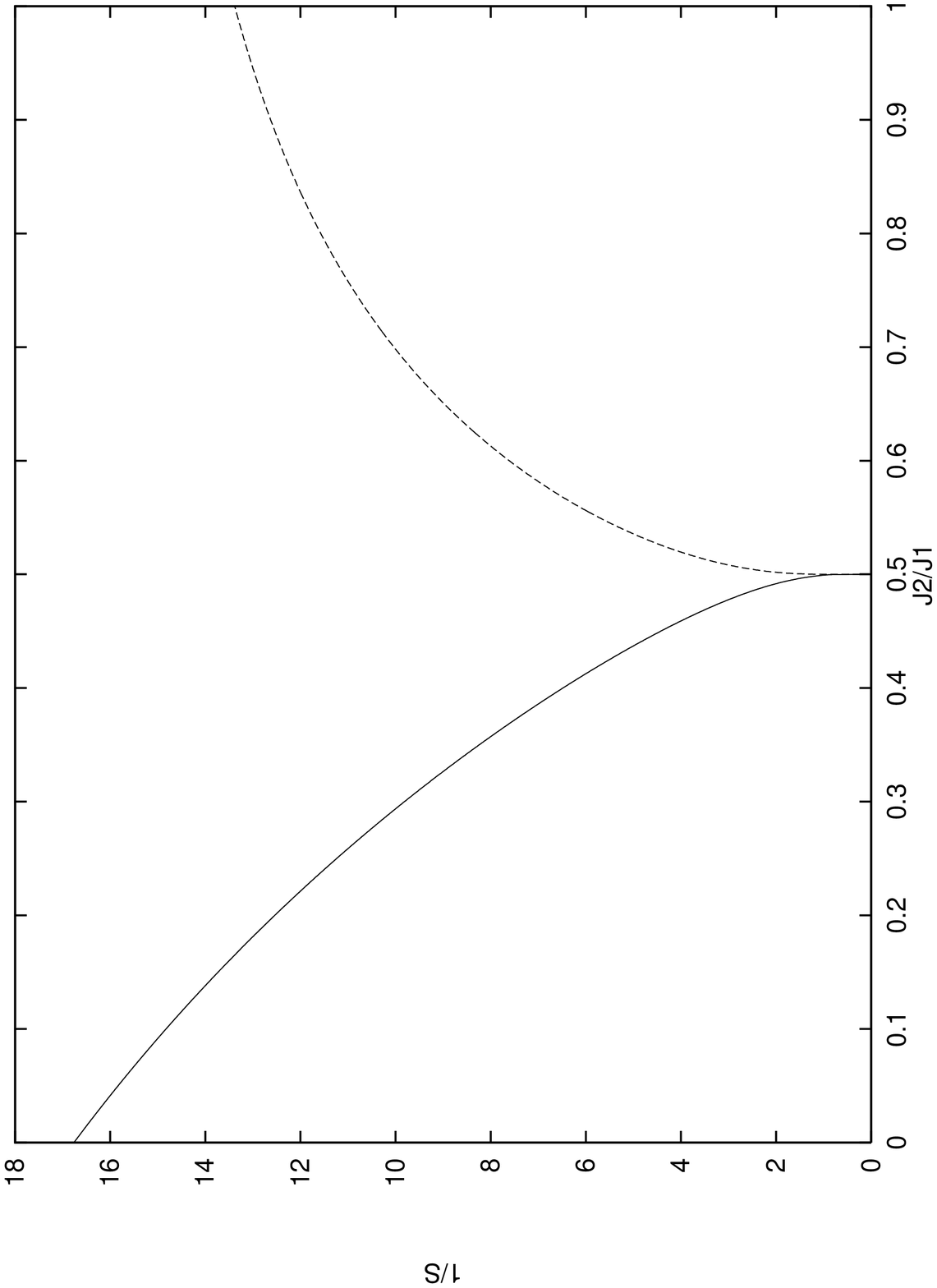,height=18cm,width=14cm}

\vskip -1cm 
FIGURE 2 
\eject

\psfig{figure=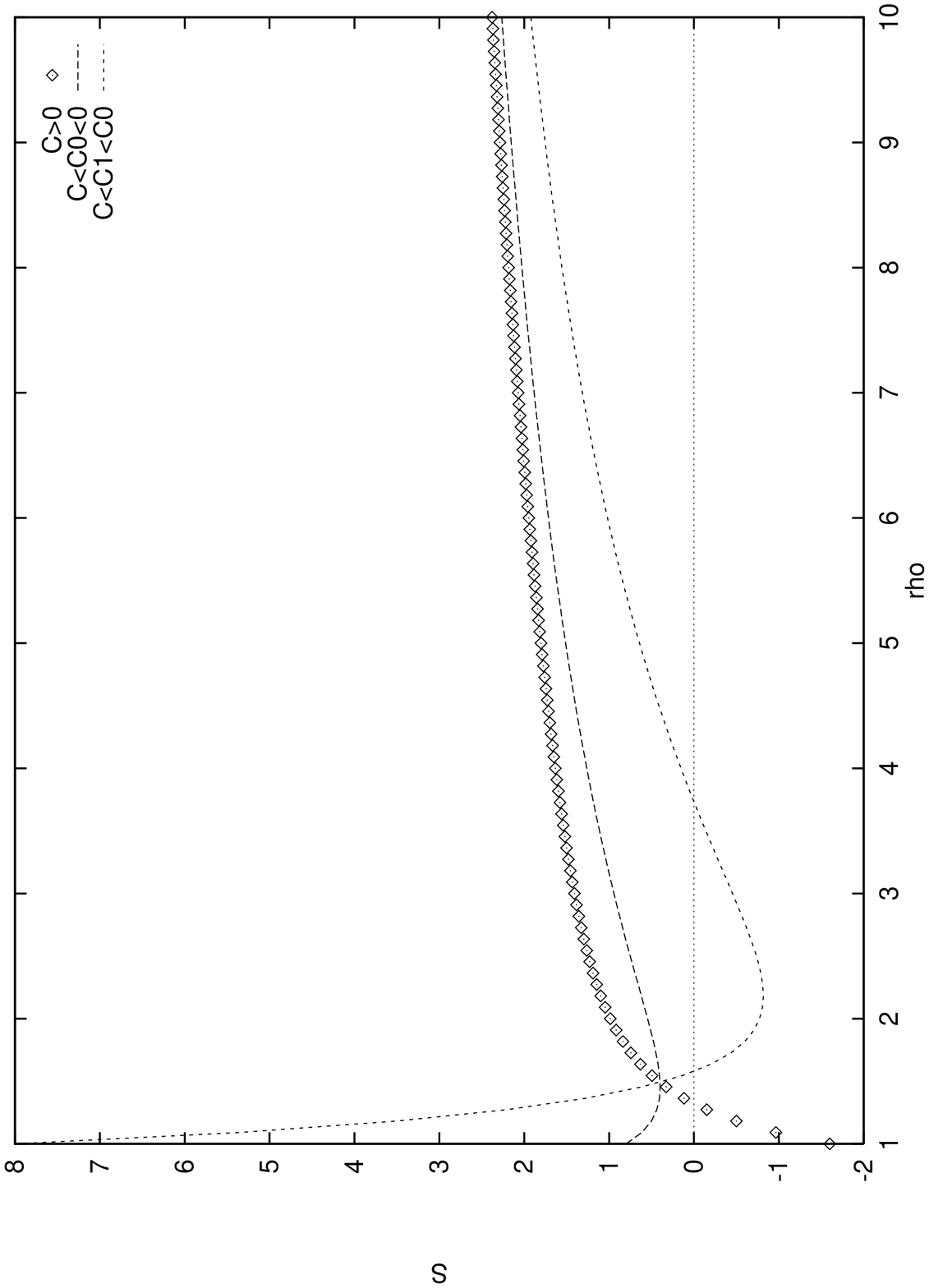,height=18cm,width=14cm}

\vskip -1cm 
FIGURE 3
\eject

FIGURE 4

\begin{figure}
\epsfxsize=14cm
$$
\epsfbox{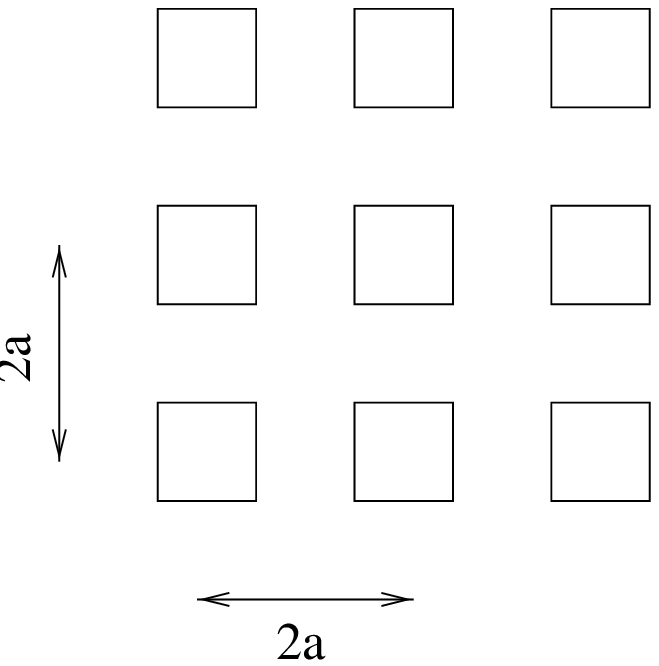}
$$
\end{figure}

\end{document}